\documentclass[amsmath,pra,twocolumn,showpacs]{revtex4}
\usepackage{graphics}

\begin{document}     

\title{Atom trap and waveguide using a two-color evanescent light field around a subwavelength-diameter optical fiber}
\author{ Fam Le Kien,$^{1,*}$ V. I. Balykin,$^{1,2}$ and K. Hakuta$^{1}$} 
\affiliation{
$^1$Department of Applied Physics and Chemistry, 
University of Electro-Communications, Chofu, Tokyo 182-8585, Japan\\
$^2$Institute of Spectroscopy, Troitsk, Moscow Region, 142092, Russia}
\date{\today}

\begin{abstract}
We suggest using a two-color evanescent light field around a \textit{subwavelength-diameter} fiber to trap and guide atoms. 
The optical fiber carries a red-detuned light and a blue-detuned light, with both modes far from resonance.  
When both input light fields are  circularly polarized, a set of 
trapping minima of the total potential in the transverse plane is formed as a ring around the fiber. This design allows confinement of atoms  to a cylindrical \textit{shell} around the fiber. 
When one or both of the input light fields are linearly polarized, the total potential has    
two local minimum points in the transverse plane. This design allows confinement of atoms to two straight \textit{lines} parallel to the fiber axis. Due to the thin thickness of the fiber, we can use far-off-resonance fields with substantially differing evanescent decay lengths to produce a net potential with  a \textit{large} depth,  a \textit{large} coherence time, and  a \textit{large} trap lifetime. For example, a 0.2-$\mu$m-radius silica fiber carrying 30 mW of
1.06-$\mu$m-wavelength  light   and 29 mW of 700-nm-wavelength  light, both fields  are circularly polarized at the input, gives for cesium atoms a trap depth of 2.9 mK, a coherence time of 32 ms, 
and a recoil-heating-limited trap lifetime of 541 s.
\end{abstract}

\maketitle

\section{Introduction}

One of the key problems of matter-wave physics is trapping and guiding neutral atoms  \cite{Nobel prizers}. Atom traps and atom waveguides can be used as tools for atom optics, atom interferometry, and atom lithography \cite{atom optics}. Atoms can be trapped and manipulated by the gradient forces of light waves \cite{dipole force}. In particular, the  evanescent light waves have been used extensively to trap and guide atoms
since they have high spatial gradients and use rigid dielectric structures like prisms and fibers to define the potential shape \cite{evanescent,b7,b4,b5}. 
An example is a hollow optical fiber with light propagating in the glass and tuned far to blue of atomic resonance   \cite{b5}.   Inside  the fiber, the evanescent wave decays exponentially away from the wall, producing a repulsive potential which guides atoms along the center axis. Alternatively, a red-detuned light in the hollow center of the fiber can also be used to guide atoms \cite{b4}. 
In several experiments  \cite{b7}, cold atoms have been trapped and guided inside a hollow fiber.

Recently,  a  method for trapping and guiding  neutral atoms 
outside a thin optical fiber has been proposed \cite{paper 1}. 
The scheme is based on the use of a subwavelength-diameter silica fiber with a red-detuned light launched into it. The light wave decays away from the fiber wall and produces an attractive potential for  neutral atoms. 
To sustain a stable trapping and guiding outside the fiber, the atoms have to be kept away from the fiber wall. This can be achieved by a centrifugal potential barrier \cite{paper 1}.  

Another way to produce a trap with a potential minimum outside a surface is to employ  
the idea of Ovchinnikov \textit{et al.} \cite{Ovchinnikov}, who proposed 
the use of two colors (i.e., red and blue detunings) and differing evanescent decay lengths to obtain both attractive and repulsive forces.
The two-color method has been considered for planar prisms \cite{Ovchinnikov}, dielectric microspheres \cite{Kimble},  free-standing channel waveguides \cite{Prentiss}, and integrated optical waveguides \cite{Julienne}. In these systems,  
the net trapping potential is small compared to the optical potentials of the red- and blue-detuned lights.
Such traps are sensitive to small field perturbations. As already pointed out by Barnett \textit{et al.} \cite{Prentiss}, such sensitivity would be greatly reduced if we could increase the difference between the evanescent decay lengths of the red- and blue-detuned lights. In addition, utilizing large detunings would also
be advantageous for coherent guiding \cite{Prentiss,Julienne}.

In this paper, we demonstrate that a \textit{subwavelength-diameter} optical fiber carrying a red-detuned light and a blue-detuned light can be used to trap and guide atoms outside  the fiber.  
Due to the thin thickness of the fiber, we can use  
far-off-resonance  lights with substantially differing evanescent decay lengths 
to produce a net potential with (1) a deep minimum, (2) a large coherence time, and (3) a large trap lifetime. We consider two schemes of input field polarization. 
In the scheme where both light fields are circularly polarized at the input, a set of 
trapping minima of the potential in the transverse plane is formed as a ring around the fiber, and
the atoms can be confined to a cylindrical \textit{shell} around the fiber. 
In the scheme where one or both of the input light fields are linearly polarized, the potential has   
two local minimum points in the transverse plane, and the atoms can be confined along two straight \textit{lines} parallel to the fiber axis. 

Before we proceed, we note that, due to recent developments in taper fiber technology, thin fibers can be produced with diameters down to 50 nm \cite{Mazur's Nature,Birks}. Therefore, a two-color trap using a subwavelength-diameter fiber is a quite realistic design.
In addition to the potential practical applications for trapping and guiding of atoms, a tapered fiber with
an intense evanescent field  can also be used as an atomic mirror \cite{Bures and Ghosh}. Generation of light with a supercontinuum spectrum in thin tapered fibers has been demonstrated \cite{Birks}. 
The evanescent waves from zero-mode metal-clad subwavelength-diameter waveguides have been used for optical observations of single-molecule dynamics \cite{zero mode}. 
Rigorous calculations for the spatial distribution, the waveguide dispersion, and the polarization orientation of the light field around a thin fiber have been reported \cite{Bures and Ghosh,Mazur's OpEx,paper 2}.
Thin fiber structures can be used as building blocks in the future atom and photonic micro and nano devices.

The paper is organized as follows. In Sec.\ \ref{sec:circ} we study the scheme where both  input light fields are  circularly polarized. In Sec.\ \ref{sec:linear} we study the scheme where one or both of the input light fields are linearly polarized. Our conclusions are given in Sec.~\ref{sec:summary}.

\section{Two-color trap with circularly polarized input lights}
\label{sec:circ}

Consider a thin  single-mode optical fiber that has a cylindrical silica core of radius $a$ and refractive index $n_1$ and  an infinite vacuum clad of refractive index $n_2=1$.
Such a fiber can be prepared using taper fiber technology. 
The essence of the technology is to heat and 
pull a single-mode optical fiber to a very thin thickness 
maintaining the taper condition to keep adiabatically 
the single-mode condition \cite{taper}. 
Due to tapering, the original core  is almost vanishing.  
Therefore, the refractive
indices that determine the guiding properties of  the tapered fiber are the refractive index of the original silica clad and the refractive index of the surrounding vacuum. 
The refractive index and the radius of the tapered silica clad will be henceforth referred to simply as the fiber refractive index $n_1$ and the fiber radius $a$, respectively. 

\subsection{Two-color optical potential}

\begin{figure}
\begin{center}
  \includegraphics{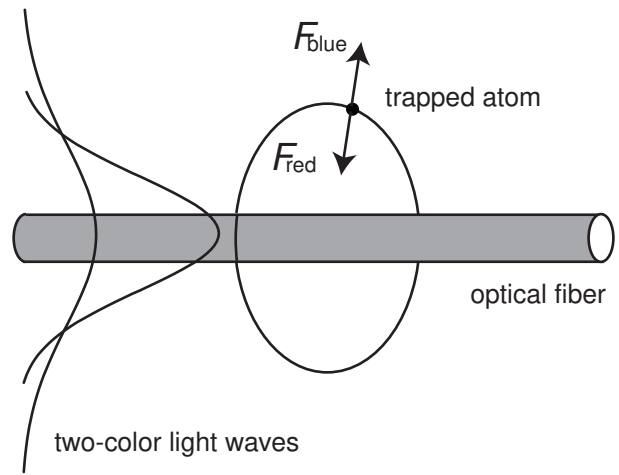}
 \end{center}
\caption{Schematic of atom trapping and guiding around an optical fiber.}
\label{fig1}
\end{figure}

Consider an atom moving  in a potential $U$ outside the fiber. 
If $U$ has a local minimum at a point outside the fiber, the atom can be trapped in the vicinity of this point. Our task is to create a potential with a trapping minimum sufficiently far from the fiber surface to make the effects of
surface interaction and heating negligible. For this purpose, we use two light fields   
in the fundamental modes 1 and 2 with differing frequencies $\omega_1$ and $\omega_2$, respectively 
(wavelengths $\lambda_1$ and $\lambda_2$, respectively, and free-space wave numbers $k_1$ and $k_2$, respectively). A schematic of our design is shown in Fig.~\ref{fig1}.
Assume that the single-mode condition \cite{fiber books} $V_i\equiv k_ia\sqrt{n_1^2(\omega_i)-n_2^2}<V_c\cong2.405$
is satisfied for both modes  ($i=1$ or 2). 
In addition, assume that the input fields are circularly polarized. 
In this case, the polarization of the transverse component of each propagating field rotates elliptically in time, 
the orbit  rotates circularly in space, and the spatial  distribution of the field intensity is cylindrically symmetric  \cite{paper 2}. 
Outside the fiber, in the cylindrical coordinates $\{r,\varphi,z\}$, 
the time-averaged intensity of the electric field in  mode $i$ is given by \cite{fiber books,paper 2}
\begin{equation}
|E_i|^2={\mathcal E}_{i}^2[K_0^2(q_ir)+ w_i K_1^2(q_ir)+f_iK_2^2(q_ir)].
\label{2}
\end{equation}
Here the notation $K_n$ stands for  the modified Bessel functions of the second kind,
$q_i$ characterizes the decay of the field outside the fiber,  $w_i$ and $f_i$ describe the deviations of the exact fundamental mode HE$_{11}$ from the approximate mode LP$_{01}$, 
and ${\mathcal E}_{i}$ determines the strength of the electric field.
The coefficients $w_i$ and $f_i$ are defined as 
$w_i= {2q_i^2}/{\beta_i^2(1-s_i)^2}$ and $f_i=(1+s_i)^2/(1-s_i)^2$, where
$s_i=({1}/{q_i^2a^2}+{1}/{h_i^2a^2})/
[{J_1'(h_ia)}/{h_iaJ_1(h_ia)}+{K_1'(q_ia)}/{q_iaK_1(q_ia)}]$.
The parameters $q_i$ and $h_i$ are related to the  longitudinal propagation constant $\beta_i$ as
$q_i=(\beta_i^2-n_2^2k_i^2)^{1/2}$ and $h_i=[n_1^2(\omega_i)k_i^2-\beta_i^2]^{1/2}$.
The parameter $\beta_i$  is determined by the eigenvalue equation for the fundamental mode at the frequency $\omega_i$ \cite{fiber books}.

We assume that the atom is in the ground state and  the fields are off resonance with the atom.
The optical potential of the atom in the field of mode $i$ is then given by \cite{Jackson} 
\begin{equation}
U_i=-\frac{1}{4}\alpha_i |E_i|^2,
\label{4}
\end{equation}
where $\alpha_i=\alpha(\omega_i)$ is the real part of the atomic polarizability at the optical frequency $\omega_i$. 
The factor $1/4$ in Eq.~(\ref{4}) results from the fact that the dipole of the atom is not a permanent dipole but is  induced by the field, giving $1/2$, and from the fact that the intensity is averaged over optical oscillations, giving another $1/2$. 

In general, the function $\alpha(\omega)$ for a ground-state atom is given by \cite{Jackson}
$\alpha(\omega)=({e^2}/{m_e})\sum_j f_{aj}(\omega_{ja}^2-\omega^2)/[(\omega_{ja}^2-\omega^2)^2+\gamma_{ja}^2\omega^2]$.
Here $e$ and $m_e$ are the electric charge and mass, respectively, of the electron, and 
$\omega_{ja}$, $f_{aj}$, and $\gamma_{ja}$ are the frequency, absorption oscillator strength, and
emission transition probability, respectively, of the spectral line $ja$. 
When we use the relation $\gamma_{ja}=f_{aj}e^2\omega_{ja}^2g_a/2\pi m_e\epsilon_0c^3g_j$,
we find
\begin{equation}
\alpha(\omega)=2\pi\epsilon_0 c^3\sum_j \frac{g_j}{g_a}
\frac{\gamma_{ja} (1-\omega^2/\omega_{ja}^2)}{(\omega_{ja}^2-\omega^2)^2+\gamma_{ja}^2\omega^2}.
\label{28}
\end{equation}
Here $g_j$ is the statistical weight of the excited level $|j\rangle$, and $g_a$ is the statistical weight of the ground-state manifold $|a\rangle$.

We note that, in the model of a two-level atom, 
the real part of the polarizability can be approximated as 
$\alpha=-{\pi\epsilon_0c^3\gamma_{ba}}/{\omega_{ba}^3\Delta}=-{d^2}/{\hbar\Delta}$.
Here $\Delta=\omega-\omega_{ba}$ is the detuning of the optical frequency $\omega$ from the atomic frequency $\omega_{ba}$ and $d$ is  the  projection of the dipole moment onto an axis. 
In deriving the above approximation for $\alpha$, it has been assumed that $\Delta$
is large compared to the linewidth $\gamma_{ba}$ but is small compared to the optical and atomic frequencies.
The corresponding approximate expression for the optical potential of the atom is $U_0=\hbar \Omega^2/\Delta$ where $\Omega=d|E|/2\hbar$ is the Rabi frequency.
When the field frequency $\omega$ is far from resonance with the atomic frequencies $\omega_{ja}$, we must take into account the multilevel structure of the atom.

Assume that the timescale of atomic motion is much slower than the beating period of the two light fields, that is, the inverse of their frequency difference. Then, the optical potentials of the two fields add up to give the net optical potential
\begin{equation}
U=U_1+U_2.
\label{3}
\end{equation}
The sign of the optical potential of each mode  is controlled by the sign of the mode detuning. 
We choose a red detuning ($\Delta_1<0$) for the field in mode 1 and a blue detuning ($\Delta_2>0$) for the field in  mode 2. 
This allows both attractive (red-detuned) and repulsive (blue-detuned) potentials to be created. 
When we substitute Eq.~(\ref{2}) into Eq.~(\ref{4}) and then the result into Eq.~(\ref{3}), we obtain
\begin{eqnarray}
U(r)&=&
G_2[K_0^2(q_2r)+ w_2 K_1^2(q_2r)+f_2K_2^2(q_2r)]
\nonumber\\&&\mbox{}
-G_1[K_0^2(q_1r)+ w_1 K_1^2(q_1r)+f_1K_2^2(q_1r)].\qquad
\label{6}
\end{eqnarray}
Here $G_1=\alpha_1 {\mathcal E}_{1}^2/4=|\alpha_1| {\mathcal E}_{1}^2/4$ and 
$G_2=-\alpha_2 {\mathcal E}_{2}^2/4= |\alpha_2| {\mathcal E}_{2}^2/4$ are positive coupling parameters. They are proportional to the powers of the corresponding light fields.
Since $\omega_1<\omega_2$, we have $q_1<q_2$, that is, the evanescent decay length $\Lambda_1=1/q_1$ of the red-detuned field is larger than the evanescent decay length $\Lambda_2=1/q_2$ of the blue-detuned field \cite{fiber books, paper 1}.
When the intensity of the blue-detuned field is large enough, the net optical potential $U$ 
is repulsive  at short range and  attractive  at long range \cite{Ovchinnikov}. Such a potential possesses a local minimum point, where the two forces cancel each other, leading to the possibility of atom trapping.
Since $U$ is independent of  $\varphi$ and $z$,  
a minimum point in the one-dimensional space $\{r\}$ corresponds to a cylindrical shell of 
minimum points in the three-dimensional space $\{r,\varphi,z\}$.

To get insight into the trapping properties of the net optical potential $U$, we perform an analytical treatment. 
To simplify this treatment, we assume that, from one side, the coefficients $w_i$ and $f_i$ are sufficiently small and, from another side, the parameters $q_ia$ are sufficiently large that the contributions of the terms containing $w_i$ and $f_i$ in Eq.~(\ref{6}) are not substantial  \cite{paper 2}. Then, we obtain the following approximate expression:
\begin{equation} 
U(r)= G_2K_0^2(q_2r)-G_1K_0^2(q_1r).
\label{6a}
\end{equation} 
We use this approximation only for our analysis but not for our numerical calculations.

We introduce the notation $r_m$ and $U_m=U(r_m)$ for the local minimum point and local minimum value of $U$, respectively. 
From $U'(r_m)=0$, we find    
\begin{equation}
F(r_m)=A, 
\label{7}
\end{equation}
where 
\begin{equation}
F(r)=\frac{K_0(q_1r)K_1(q_1r)}{K_0(q_2r)K_1(q_2r)} 
\label{8}
\end{equation}
and
\begin{equation}
A=\frac{q_2G_2}{q_1G_1}.
\label{9}
\end{equation}
Since  $q_1<q_2$, the function $F(r)$ is a monotonically increasing function of $r$. Hence, Eq.~(\ref{7}) has a solution $r_m>a$ if $A> F(a)$, that is, if
\begin{equation}
\frac{G_2}{G_1}> \frac{q_1K_0(q_1a)K_1(q_1a)}{q_2K_0(q_2a)K_1(q_2a)}.
\label{10}
\end{equation}
Condition (\ref{10}) means that there exists a trapping minimum outside the fiber if
the intensity of the blue-detuned light is large enough or if the intensity of the red-detuned light is small enough (but not zero).
Note that, when the power of the  blue-detuned light increases or the power of the  red-detuned light decreases, the depth $U_D\equiv-U_m=|U_m|$ of the trapping potential decreases and
the local minimum point $r_m$  is shifted farther away from the fiber surface.
Oppositely, when the power of the  blue-detuned light decreases or the power of the  red-detuned light increases, the potential depth increases and
the local minimum point is shifted towards the fiber surface.

The atom can be prevented from coming too close to the fiber surface before being loaded into the trap if 
the net optical potential  achieves a non-negative value at the fiber surface, that is, if  $U(a)\geq0$. This requirement will be fulfilled if
\begin{eqnarray}
\frac{G_2}{G_1}\geq\frac{K_0^2(q_1a)}{K_0^2(q_2a)}.
\label{11}
\end{eqnarray}
When condition (\ref{11}) is satisfied, condition (\ref{10}) is also satisfied.
To maximize the depth of the trapping minimum, we optimize the powers of the two field in such a way that  
condition (\ref{11}) reduces to an equality, namely,   
\begin{eqnarray}
\frac{G_2}{G_1}=\frac{K_0^2(q_1a)}{K_0^2(q_2a)}.
\label{11a}
\end{eqnarray}
Then, Eq.~ (\ref{6a}) yields  
\begin{eqnarray}
U(r)&=&G_0\left[\frac{K_0^2(q_2r)}{K_0^2(q_2a)}-\frac{K_0^2(q_1r)}{K_0^2(q_1a)}\right],
\end{eqnarray}
where $G_0=G_1 {K_0^2(q_1a)}=G_2{K_0^2(q_2a)}$.

We analyze the effects of the evanescent decay parameters $q_1$ and $q_2$ on the trapping potential under condition (\ref{11a}).
For this purpose, we fix  $q_1$ and $G_1$ (or $q_2$ and $G_2$) and increase $q_2$ and $G_2$  
(or decrease $q_1$ and $G_1$) in such a way that equality (\ref{11a}) is kept.
This leads to a reduction in $U$ and, consequently,
to a decrease in the minimum value $U_m$, that is, to an increase in the depth $U_D=-U_m$ of the
trapping potential. Such variations also make the minimum point of the potential  closer to the fiber surface.
Note that, when the fiber radius is small, the evanescent decay length $\Lambda$ is a fast increasing function of the light wavelength $\lambda$ \cite{paper 1}.
Therefore, using a thin fiber, we can obtain a large ratio of $q_2$ to $q_1$ by choosing a small  $\lambda_2$ and/or a large $\lambda_1$. 
Thus, a reduction in the blue-detuned light wavelength $\lambda_2$ or an increase in the red-detuned light wavelength $\lambda_1$ allows us to increase the potential depth provided
the power of the light field $E_2$ or $E_1$, respectively, can be optimized appropriately to satisfy condition (\ref{11a}).
However,  $\lambda_2$ must not be too small because
it is limited by the single-mode condition. In addition, $\lambda_1$ must not be too large because a larger wavelength $\lambda_1$ leads to a larger penetration length $\Lambda_1$ and, therefore, to a smaller potential magnitude for the same power of light. 
Furthermore,  the detunings $\Delta_i$ of the light fields from the atomic resonance frequency must not be too large because a larger detuning leads to a smaller
polarizability and, consequently,  to a smaller potential magnitude for the same power of light.  

In general, the polarizability of an atom is a complex characteristic. The imaginary part of the polarizability is given by \cite{Jackson}  
\begin{equation}
\kappa(\omega)
=2\pi\epsilon_0 c^3\sum_j \frac{g_j}{g_a}
\frac{\gamma_{ja}^2\omega/\omega_{ja}^2}{(\omega_{ja}^2-\omega^2)^2+\gamma_{ja}^2\omega^2}.
\label{16}
\end{equation}
It is responsible for spontaneous scattering. The rate of spontaneous scattering caused by a single light field $E_i$ is given by 
\begin{equation}
\Gamma_i^{\mathrm{(sc)}}=\frac{1}{4\hbar}\kappa_i |E_i|^2,
\label{17}
\end{equation}
where $\kappa_i=\kappa(\omega_i)$.
Spontaneous scattering limits the coherence time of the trap. For atoms spending time close to the minimum $r_m$ 
of the two-color potential $U$, the net scattering rate is 
\begin{equation}
\Gamma_{\mathrm{sc}}=\Gamma_1^{\mathrm{(sc)}}(r_m)+\Gamma_2^{\mathrm{(sc)}}(r_m) 
\label{17a}
\end{equation}
and the characteristic coherence time is \cite{Prentiss,Julienne}
\begin{equation}
\tau_{\mathrm{coh}}=\frac{1}{\Gamma_{\mathrm{sc}}}. 
\label{17b}
\end{equation}
Every scattered photon imparts a recoil energy $\theta_i^{\mathrm{(rec)}}=(\hbar k_i)^2/2M$ to the atom, where $M$ is the mass of the atom. Therefore,
the absorption of mode photons and emission of other photons result in a loss of atoms from the trapping potential. For a trap depth $U_D$, the quantity  \cite{Julienne}
\begin{equation}
\tau_{\mathrm{trap}}=\frac{U_D}{2\sum_i\theta_i^{\mathrm{(rec)}}\Gamma_i^{\mathrm{(sc)}}(r_m)}
\label{17c}
\end{equation}
characterizes the trap lifetime due to recoil heating. 
When the light field frequencies are near to the atomic resonances, the scattering rates are large and, therefore,
the coherence time and the trap lifetime  are small. To produce a trap with a large coherence time and a large trap lifetime, we must use far-off-resonance fields.

\subsection{van der Waals potential}

An atom near the surface of a medium undergoes a van der Waals force.  
The van der Waals potential of an atom near the surface of a cylindrical dielectric rod 
is given by \cite{Baudon}
\begin{eqnarray}
V(r)&=&\frac{\hbar}{4\pi^3\epsilon_0}\sum_{n=-\infty}^{\infty}\int_0^{\infty} dk\,
[k^2K_n'^2(kr)\nonumber\\
&&\mbox{}+(k^2+n^2/r^2)K_n^2(kr)]\int_0^\infty d\xi\, \alpha(i\xi)G_n(i\xi),
\nonumber\\
\label{29}
\end{eqnarray}
where
\begin{equation}
G_n(\omega)=\frac{[\epsilon(\omega)-\epsilon_0]I_n(ka)I_n'(ka)}{\epsilon_0I_n(ka)K_n'(ka)-\epsilon(\omega)I_n'(ka)K_n(ka)}.
\label{30}
\end{equation}
Here $\epsilon(\omega)$ is the dynamical dielectric function and $I_n$ is the modified Bessel function of the first kind.

We calculate the van der Waals potential of a ground-state cesium atom near a silica fiber surface. 
The dynamical dielectric function of silica is given by \cite{Si}
$\frac{\epsilon(\omega)}{\epsilon_0}=1+\frac{0.6961663\,\lambda^2}{\lambda^2-0.0684043^2}
+\frac{0.4079426\,\lambda^2}{\lambda^2-0.1162414^2}
+\frac{0.8974794\,\lambda^2}{\lambda^2-9.896161^2}$,
where $\lambda$ is in the units of $\mu$m. 
To calculate the integral (\ref{29}), we use the approximate expression 
$\alpha(\omega)=2\pi\epsilon_0 c^3\sum_j 
g_j\gamma_{ja}/[g_a\omega_{ja}^2(\omega_{ja}^2-\omega^2)]$.
This approximation is justified because the resonant frequencies of the ground-state cesium atom  are substantially different from the resonant frequencies of silica.
We take into account four dominant lines of the atom, namely,
$\lambda_{1a}=852.113$ nm, $\lambda_{2a}=894.347$ nm, $\lambda_{3a}=455.528$ nm, 
and $\lambda_{4a}=459.317$ nm, see \cite{Cs}. The emission transition probabilities of these lines are
$\gamma_{1a}=3.276\times 10^7$ s$^{-1}$, $\gamma_{2a}=2.87\times 10^7$ s$^{-1}$, 
$\gamma_{3a}=1.88\times 10^6$ s$^{-1}$, and $\gamma_{4a}=8\times 10^5$ s$^{-1}$.
The statistical weights of the four corresponding upper states are $g_1=4$, $g_2=2$, $g_3=4$, and $g_4=2$. 
The statistical weight of the ground state is $g_a=2$.

\begin{figure}
\begin{center}
  \includegraphics{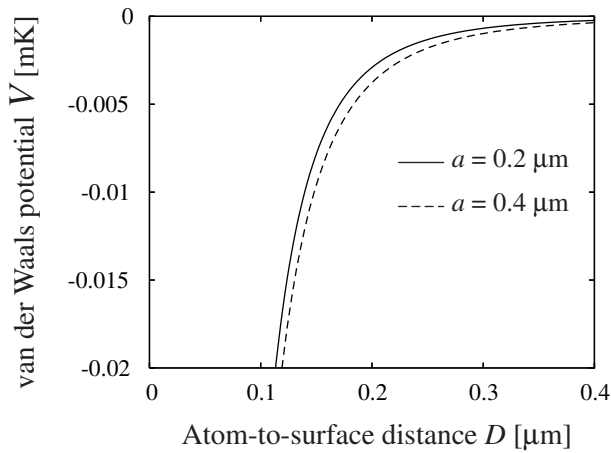}
 \end{center}
\caption{van der Waals potential $V$ of a ground-state cesium atom near a thin cylindrical silica fiber.
}
\label{V}
\end{figure}

In Fig.~\ref{V}, we plot  the van der Waals potential $V$ of a ground-state cesium atom near a cylindrical silica fiber as a function of the atom-to-surface distance $D=r-a$.  
The comparison between the solid line ($a=0.2$ $\mu$m) and the dashed line ($a=0.4$ $\mu$m)
shows that a smaller fiber radius $a$ leads to a smaller magnitude and a less steep slope of the van der Waals potential $V$. 

\begin{figure}
\begin{center}
  \includegraphics{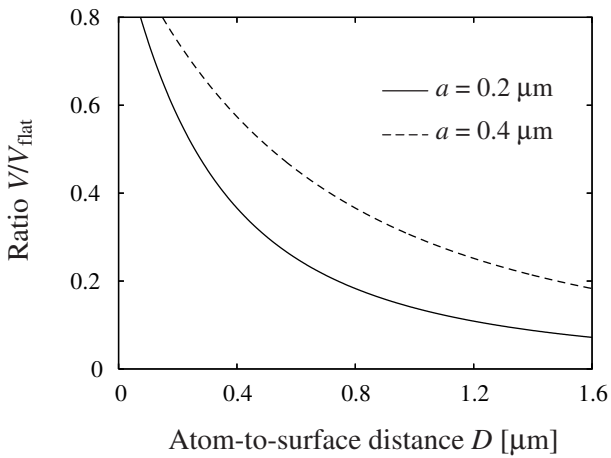}
 \end{center}
\caption{Ratio between the van der Waals potentials $V$ and $V_{\mathrm{flat}}$
of a ground-state cesium atom near a thin cylindrical silica fiber and  
a flat-surface bulk silica dielectric, respectively. 
}
\label{Vflat}
\end{figure}

When the fiber radius $a$ is small compared to the atom-to-surface distance $D$, the van der Waals potential $V$ is, in general, different from the flat-surface bulk-medium van der Waals potential 
$V_{\mathrm{flat}}=-C_3/D^3$. 
Here the coefficient $C_3$ is determined by \cite{McLachlan}
$C_3=({\hbar}/{16\pi^2\epsilon_0})
\int_0^\infty d\xi\, \alpha(i\xi)[\epsilon(i\xi)-\epsilon_0]/[\epsilon(i\xi)+\epsilon_0]$.
For cesium atoms and flat silica surfaces, this coefficient is estimated to be $C_3\cong5.6\times10^{-49}$ $\mathrm{J\, m}^3\cong4.1\times 10^{-5}$ $\mathrm{mK}\,\mu \mathrm{m}^3$.
Figure \ref{Vflat} illustrates the difference between the fiber-surface potential $V$ 
and the flat-surface potential $V_{\mathrm{flat}}$. 
The figure shows that $V/V_{\mathrm{flat}}<1$, that is, the magnitude of $V$  is smaller  than the magnitude of $V_{\mathrm{flat}}$. 
When the atom-to-surface distance $D$ tends to zero, the ratio $V/V_{\mathrm{flat}}$ tends to unity, that is, the two potentials $V$ and $V_{\mathrm{flat}}$ tend to become the same.
When $D$ increases, the ratio $V/V_{\mathrm{flat}}$ reduces, that is, the relative difference between $V$ and $V_{\mathrm{flat}}$ increases.

\subsection{Total potential}

The total  potential $U_{\mathrm{tot}}$ of the atom is the sum of the net optical potential $U$ and the van der Waals potential $V$, i.e.,  
\begin{equation}
U_{\mathrm{tot}}=U+V.
\label{1}
\end{equation}
We use Eqs.~(\ref{6}), (\ref{29}), and (\ref{1}) to calculate  the total potential $U_{\mathrm{tot}}$ 
of a ground-state cesium atom outside a thin cylindrical silica fiber  with two circularly polarized input light fields. The ground-state cesium atom has  two strong transitions, at 852 nm ($D_2$ line) and 894 nm ($D_1$ line).
To trap the atom, we use  red- and blue-detuned lights with wavelengths $\lambda_1=1.06$ $\mu$m and $\lambda_2=700$ nm, respectively. The detunings of the lights from the dominant $D_2$ line of the atom are $\Delta_1/2\pi=-69$ THz and $\Delta_2/2\pi=76$ THz.
The fiber radius is $a=0.2$ $\mu$m. This radius is  small enough 
to create a large ratio of $q_2$ to $q_1$ and to reduce the effect of the van der Waals force \cite{paper 1}. 
For the above parameters, we find $q_1a\cong 0.2438$ and $q_2a\cong 0.9686$.
The corresponding evanescent decay lengths are $\Lambda_1\cong 0.8$ $\mu$m  and $\Lambda_2\cong 0.2$ $\mu$m.
The relative difference between the decay lengths is measured by the parameter $\alpha_\Lambda=(\Lambda_1-\Lambda_2)/\Lambda_2\cong 3$. The obtained value of this parameter is much larger than the characteristic value $\alpha_\Lambda\cong0.47$ estimated for the case of channel guides with the TE and TM modes \cite{Prentiss}. 

\begin{figure}
\begin{center}
  \includegraphics{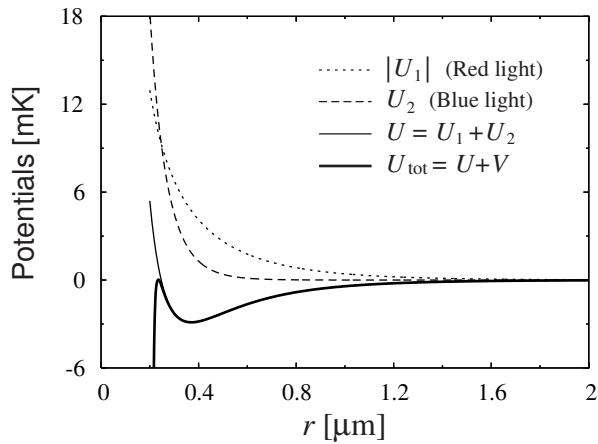}
 \end{center}
\caption{Contributions to the total trapping potential $U_{\mathrm{tot}}$ of a ground-state cesium atom 
outside a vacuum-clad subwavelength silica fiber with two circularly polarized input light fields.  The radius of the fiber is $a=0.2$ $\mu$m. 
The absolute value of the red-detuned component  $U_1$
(dotted line) subtracts from the blue-detuned component $U_2$ (dashed line) to give the net optical potential $U$ (thin solid line). This is modified by the van der Waals surface interaction, giving the  total potential $U_{\mathrm{tot}}$ (thick solid line). The light wavelengths  are $\lambda_1=1.06$ $\mu$m  and $\lambda_2=700$ nm. The light powers are  $P_1=30$ mW and $P_2=29$ mW.} 
\label{contr}
\end{figure}

In Fig.~\ref{contr}, we plot the contributions to the total trapping potential $U_{\mathrm{tot}}$ of the atom. 
The figure shows that the evanescent decay length of the red-detuned light is substantially larger than that of the blue-detuned light. The net optical potential $U$ has a local  minimum value $U_m\cong - 2.9$ mK at $r_m\cong 0.37$ $\mu$m, well outside the fiber. The van der Waals potential is only about $-5$ $\mu$K at the distance $r_m-a\cong 0.17$ $\mu$m from the fiber surface and makes just a small contribution in the vicinity of the trapping minimum. It is interesting to note that the depth $U_D\cong 2.9$ mK of the net optical potential is comparable to the red-detuned potential and is larger than the blue-detuned potential
at the trapping point. These characteristics of the two-color fiber scheme are much better than those of the two-color integrated-waveguide scheme, where the red- and blue-detuned potentials are large but the net potential is small \cite{Julienne}.

We estimate some critical trapping parameters for the case of Fig.~\ref{contr}. We find that
the rates of scattering due to the trapping fields at the potential minimum are  $\Gamma_1^{\mathrm{(sc)}}\cong 22.39$ s$^{-1}$ and $\Gamma_2^{\mathrm{(sc)}}\cong 8.46$ s$^{-1}$. Accordingly,
the net scattering rate and the coherence time are  $\Gamma_{\mathrm{sc}}\cong 30.85$ s$^{-1}$  and $\tau_{\mathrm{coh}}\cong 32$ ms, respectively. 
Since the recoil energies due 
to single red- and blue-detuned photons are $\theta_1^{\mathrm{(rec)}}\cong 0.06$ $\mu$K and $\theta_2^{\mathrm{(rec)}}\cong 0.15$ $\mu$K, respectively, the trapping lifetime due to recoil heating is estimated to be $\tau_{\mathrm{trap}}\cong541$ s.
At the trapping minimum, the radial oscillation frequency is $\omega_r/2\pi\cong 492$ KHz, giving 
an estimate of about 23.6 $\mu$K for the energy of the atomic mode spacing.

\begin{figure}
\begin{center}
  \includegraphics{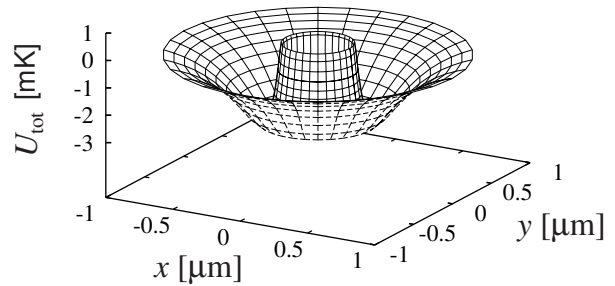}
 \end{center}
\caption{
Transverse-plane profile of the total potential $U_{\mathrm{tot}}$ produced by lights that are circularly polarized at the input. All the parameters are the same as for Fig.~\ref{contr}.
} 
\label{circ3D}
\end{figure}

Due to the circular polarization of the input fields,  $U_{\mathrm{tot}}$ is cylindrically symmetric.
In Fig.~\ref{circ3D}, we plot the spatial profile of $U_{\mathrm{tot}}$ in the fiber transverse plane.
The figure shows that $U_{\mathrm{tot}}$ is independent of the
azimuthal angle $\varphi$ and has a set of deep minima  formed as a ring surrounding
the fiber. Since the atom can move freely along the direction $z$ of the fiber axis, $U_{\mathrm{tot}}$ is also independent of  $z$. Therefore, the ring of the potential minima in the transverse plane results in a cylindrical shell of the minima of the  trapping potential in the three-dimensional space.

\begin{figure}
\begin{center}
  \includegraphics{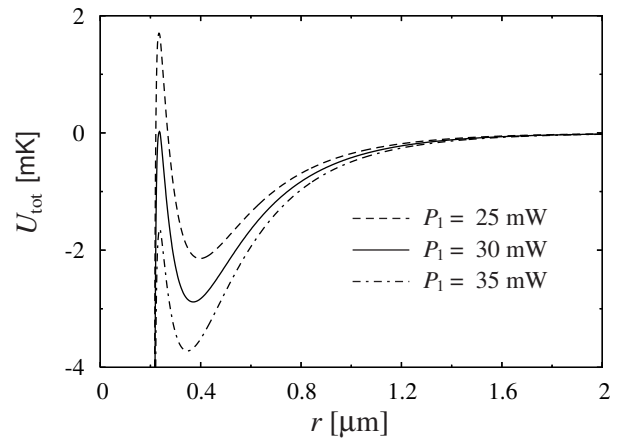}
 \end{center}
\caption{Effect of the power $P_1$ of the red-detuned light on the total  potential $U_{\mathrm{tot}}$. 
The power  of the blue-detuned light is fixed at $P_2=29$ mW. 
All the other parameters are the same as for Fig.~\ref{contr}.
} 
\label{powred}
\end{figure}

Figure \ref{powred} illustrates the effect of  the power $P_1$ of the red-detuned light on the total potential $U_{\mathrm{tot}}$. As seen, an increase in $P_1$ leads to an increase in the depth $U_D$ of the trapping potential and to a shift of the local minimum  point $r_m$ towards the fiber surface. However, the increase in $P_1$ also reduces the height of the repulsive wall in the region of $r<r_m$. 

\begin{figure}
\begin{center}
  \includegraphics{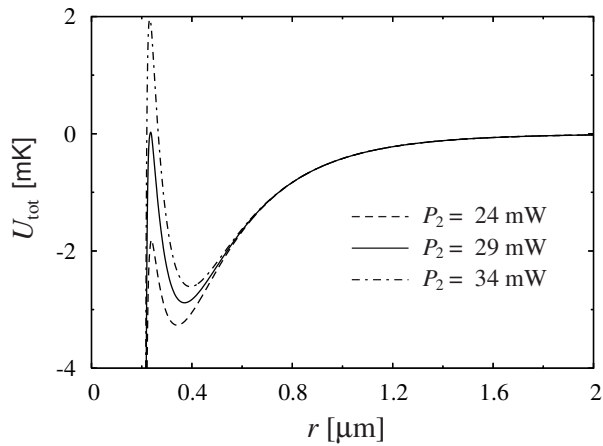}
 \end{center}
\caption{
Effect of the power $P_2$ of the blue-detuned light on the total  potential $U_{\mathrm{tot}}$. 
The power of the red-detuned light is fixed at $P_1=30$ mW. 
All the other parameters are the same as for Fig.~\ref{contr}.
} 
\label{powblue}
\end{figure}

Figure \ref{powblue} illustrates the effect of  the power $P_2$ of the blue-detuned light on the total potential $U_{\mathrm{tot}}$. As seen, an increase in $P_2$ leads to a decrease in the depth $U_D$ of the trapping potential and to a shift of the local minimum  point $r_m$ farther away from the fiber surface. Meanwhile, the height of the repulsive wall in the region of $r<r_m$ is increased. 
The comparison between Figs.~\ref{powred} and \ref{powblue} show that the powers $P_1$ and $P_2$ of the red- and blue-detuned lights, respectively, have opposite effects on the total potential of the atom. Therefore, to produce a potential with a deep trapping minimum outside the fiber and with a high repulsive wall in the region of $r<r_m$, the ratio between $P_1$ and $P_2$ must be optimized appropriately.

\begin{figure}
\begin{center}
  \includegraphics{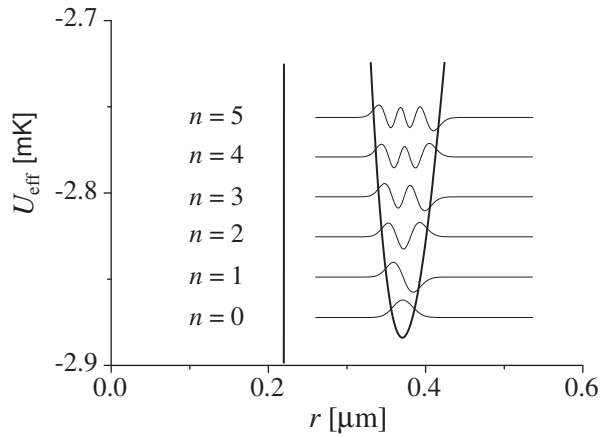}
 \end{center}
\caption{Bound states for the first six levels ($n=0$, 1, 2, 3, 4, and 5) of the radial motion of a cesium atom in the  effective potential $U_{\mathrm{eff}}=U_{\mathrm{tot}}+\hbar^2(m^2-1/4)/2M r^2$.
The rotational quantum number is $m=0$.
All the other parameters are the same as for Fig.~\ref{contr}.
} 
\label{eigen}
\end{figure}

Due to the cylindrical symmetry of the total potential $U_{\mathrm{tot}}$, the component $L_z$ of the angular momentum of the atom is conserved. 
In the eigenstate problem, we have $L_z=\hbar m$, where $m$ is an integer, called the rotational quantum number.
The centrifugal potential of the atom is given by $U_{\mathrm{cf}}={\hbar^2(m^2-1/4)}/{2M r^2}$. The  radial motion of the atom can be treated as the one-dimensional motion of a particle in the effective potential $U_{\mathrm{eff}}=U_{\mathrm{tot}}+U_{\mathrm{cf}}$. In Fig.~\ref{eigen},
we plot the eigenfunctions for the first six levels  of the radial motion of the atom in the  effective potential $U_{\mathrm{eff}}$ with the rotational quantum number  $m=0$. The energy of the ground state
is $E_0\cong-2.872$ mK. The spacing between the  the energies of the ground state and the first excited state is  roughly $23$ $\mu$K. The characteristic size of the ground state is $\Delta r\cong 8.8$ nm.

\section{Two-color trap with linearly polarized input lights}
\label{sec:linear}

The formation of a set of trapping minima as a ring in the fiber transverse plane (or as a shell in the three-dimensional space), shown in the previous section, is due to the cylindrical symmetry of  all of the components of the total potential.
In order to confine atoms to vicinities of single local points in the fiber transverse plane, we need to break the symmetry.  An asymmetric optical potential can be obtained when one or both trapping light fields  are linearly polarized at the input. 
Indeed, it has been shown that, due to 
the thin thickness of the fiber and the high  contrast between the refractive indices of the silica core and the vacuum clad, the intensity distribution of the field in a fundamental mode with quasi-linear polarization
strongly depends on the azimuthal angle, especially in the  vicinity of the fiber surface \cite{paper 2}.

Regarding the properties of the trapping potential, there is no substantial qualitative difference between the case where both input fields are linearly polarized and the case where one is linearly polarized and the other one is circularly polarized. Therefore, we consider here, as an example, the case where both input fields are linearly polarized, along the same direction, namely, the $x$ direction. In this case, the polarization of each light field propagating along the fiber is quasi-linear and the spatial distribution of the field intensity is not cylindrically symmetric \cite{paper 2}. 
Outside the fiber, the time-averaged intensity of the electric field in  mode $i$ is given by \cite{fiber books,paper 2}
\begin{eqnarray}
|E_i|^2&=&\mathcal{E}_i^2\{K_0^2(q_ir)+w_i K_1^2(q_ir)
+f_iK_2^2(q_ir)
\nonumber\\&&\mbox{}
+ [w_i K_1^2(q_ir)+\xi_{i}K_0(q_ir)K_2(q_ir)]\cos2\varphi\}.
\nonumber\\
\label{12}
\end{eqnarray}
Here $\xi_i=2{(1+s_i)}/{(1-s_i)}$. 
We use the above field intensity distributions and Eq.~(\ref{4}) to calculate the optical potentials $U_1$ and $U_2$ of an atom in the red- and blue-detuned fields.
The net optical potential $U=U_1+U_2$ is then found to be
\begin{eqnarray}
U(r,\varphi)&=&
G_2\{K_0^2(q_2r)+ w_2 K_1^2(q_2r)+f_2K_2^2(q_2r)
\nonumber\\&&\mbox{}
+ [w_2 K_1^2(q_2r)+\xi_{2}K_0(q_2r)K_2(q_2r)]\cos2\varphi\}
\nonumber\\&&\mbox{}
-G_1\{K_0^2(q_1r)+ w_1 K_1^2(q_1r)+f_1K_2^2(q_1r)
\nonumber\\&&\mbox{}
+ [w_1 K_1^2(q_1r)+\xi_{1}K_0(q_1r)K_2(q_1r)]\cos2\varphi\}.
\nonumber\\
\label{13}
\end{eqnarray}
Unlike the potential (\ref{6}),  the potential (\ref{13}) contains additional terms that vary with the azimuthal angle $\varphi$. Due to these terms, the
potential (\ref{13}) is not cylindrically symmetric. Meanwhile, when the intensity of the blue-detuned field is large enough, the potential (\ref{13}) is, like the potential (\ref{6}), repulsive  at short range and  attractive  at long range.  For a fixed angle $\varphi$, the potential $U(r,\varphi)$ as a function of $r$ must have a local minimum value $U_m(\varphi)$ achieved at a point $r_m(\varphi)$ where the repulsive and attractive forces cancel each other. Due the cylindrical asymmetry of the potential,  $U_m(\varphi)$ is not constant in $\varphi$.
Consequently, $U_m(\varphi)$ must have a global minimum value $U_g=\min{U_m(\varphi)}$, which is achieved at a finite   number of angles $\varphi_m$. Thus the potential $U(r,\varphi)$ must have a global minimum value $U_g$ achieved at a finite number of  single points $(r_m(\varphi_m),\varphi_m)$ in the fiber transverse plane.
Due to the axial  symmetry, we have $U(r,\varphi)=U(r,\pi+\varphi)$. Therefore, there must be two   minimum points  for $U(r,\varphi)$ in the transverse plane. Since the atom can move freely along the direction $z$ of the fiber axis,  the two  minimum points in the transverse plane in fact correspond to two straight lines for the minima of the  trapping potential in the three-dimensional space. These potential minimum lines can be used to trap and guide atoms along the fiber.

\begin{figure}
\begin{center}
  \includegraphics{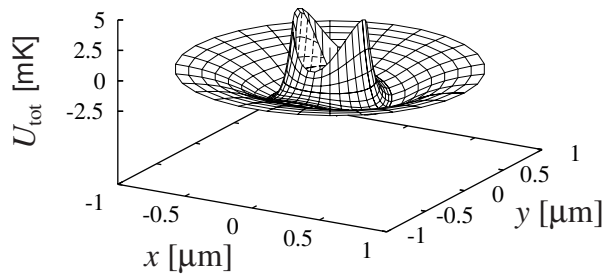}
 \end{center}
\caption{
Transverse-plane profile of the total potential $U_{\mathrm{tot}}$ produced by the lights that are linearly polarized at the input.  
The light powers are  $P_1=30$ mW and $P_2=35$ mW. 
All the other parameters are the same as for Fig.~\ref{contr}.
} 
\label{linear3D}
\end{figure}

We use Eq.~(\ref{13}) as well as Eqs.~(\ref{29}) and (\ref{1}) 
to calculate the total potential  $U_{\mathrm{tot}}$
of a ground-state cesium atom outside a thin cylindrical silica fiber with two linearly polarized input light fields. 
We plot in Fig.~\ref{linear3D} the spatial profile of $U_{\mathrm{tot}}$ in the fiber transverse plane.
The figure shows that $U_{\mathrm{tot}}$ varies substantially with the
azimuthal angle $\varphi$, especially in the vicinity of the fiber surface. The minimum points in the radial dependence
of $U_{\mathrm{tot}}$ form  a ring surrounding the fiber. Unlike the case of circular polarization, see Fig.~\ref{circ3D},
the ring at the bottom of Fig.~\ref{linear3D} is not flat with respect to $\varphi$. This ring has two deepest points, located symmetrically on the $x$ axis (at $\varphi=0$ and $\varphi=\pi$). These points correspond to two straight lines,  parallel to the fiber axis, for the trapping minima in the three-dimensional space.

\begin{figure}
\begin{center}
  \includegraphics{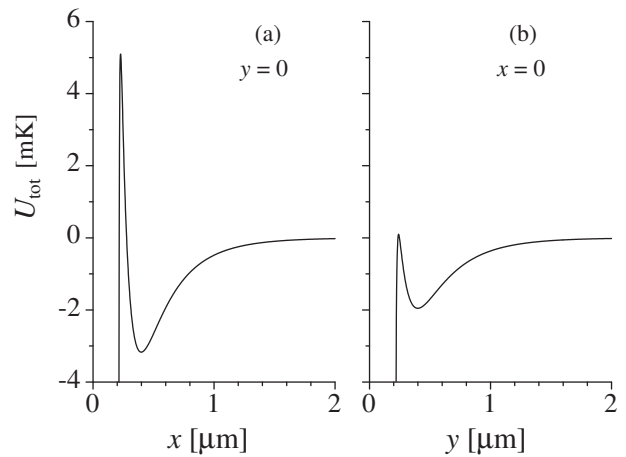}
 \end{center}
\caption{
Total potential $U_{\mathrm{tot}}$ as a function of $x$ at $y=0$ (a) and as a function of $y$ at $x=0$ (b). 
All the  parameters are the same as for Fig.~\ref{linear3D}.
} 
\label{xy}
\end{figure}

To get a closer look at the spatial distributions of the potential along different radial  directions, we replot in Fig.~\ref{xy} the potential $U_{\mathrm{tot}}$ of Fig.~\ref{linear3D} 
as a function of $x$ at $y=0$ (i.e., as a function of $r$ at $\varphi=0$)  
and as a function of $y$ at $x=0$ (i.e., as a function of $r$ at $\varphi=\pi/2$).
Figures \ref{xy}(a) and  \ref{xy}(b) show that, for $P_1=30$ mW and $P_2=35$ mW, 
the depths of the potential minima for the $x$ and $y$ directions are
about $-3.2$ mK and $-2$ mK, respectively. Thus the depth of the potential minimum for the $x$ direction 
is  deeper than that for the $y$ direction. Consequently, the local minimum for the $x$ direction is the local minimum for the whole transverse plane. The difference between the depths of the potential minima for the $x$ and $y$ directions is about 1.2 mK. Note that the radial distances $x_m=r_m(0)$ and $y_m=r_m(\pi/2)$ of the local minima for the $x$ and $y$ directions are almost the same, namely, $r_m(\varphi)\cong 0.4$ $\mu$m.

\begin{figure}
\begin{center}
  \includegraphics{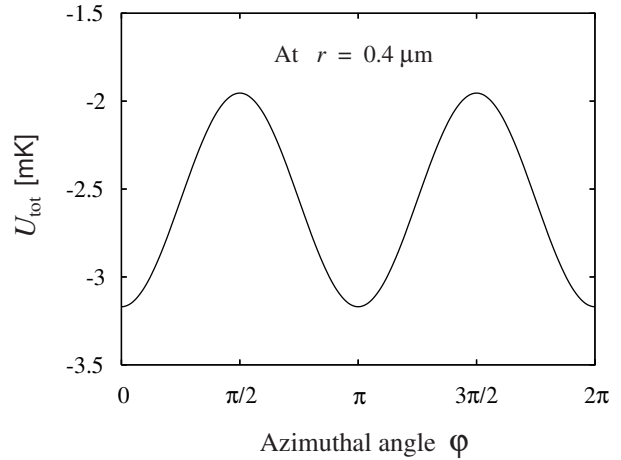}
 \end{center}
\caption{
Azimuthal dependence of the total potential $U_{\mathrm{tot}}$ for $r=0.4$ $\mu$m.  
All the parameters are the same as for Fig.~\ref{linear3D}.} 
\label{phi}
\end{figure}

To get insight into the variations of $U_{\mathrm{tot}}$ along the bottom ring  $r=r_m(\varphi)\cong 0.4$ $\mu$m, we plot this potential in Fig.~\ref{phi} as a function of $\varphi$ for a fixed radial distance $r=0.4$ $\mu$m.  As seen, along the bottom ring, $U_{\mathrm{tot}}$ achieves its  minimum value at two  angles,  $\varphi=0$ and $\varphi=\pi$. The difference between the maximum and minimum values of $U_{\mathrm{tot}}$  along the bottom ring is about 1.2 mK.

We estimate some important parameters of the trap at the minimum point $(r=r_m,\varphi=0)$ or $(r=r_m,\varphi=\pi)$ in the situation of Fig.~\ref{linear3D}. We find that the rates of scattering due to the light  fields at the trapping minimum are  $\Gamma_1^{\mathrm{(sc)}}\cong 25.87$ s$^{-1}$ and $\Gamma_2^{\mathrm{(sc)}}\cong 10.49$ s$^{-1}$. Accordingly,
the net scattering rate and the coherence time are  $\Gamma_{\mathrm{sc}}\cong 36.36$ s$^{-1}$  and $\tau_{\mathrm{coh}}\cong 27.5$ ms, respectively. 
The trapping lifetime due to recoil heating is estimated to be $\tau_{\mathrm{trap}}\cong500$ s.
When we fit the bottom of the potential around the minimum point by a 2D harmonic potential,
we find that the transverse oscillation frequencies are $\omega_r/2\pi\cong 533$ KHz and
$\omega_\varphi/2\pi\cong 156$ KHz. These frequencies provide
a ground-state localization of  $l_r\cong8.4$ nm and $l_\varphi\cong15.6$ nm.

\section{Conclusions}
\label{sec:summary}

We have demonstrated that a \textit{subwavelength-diameter} optical fiber carrying a red-detuned light and a blue-detuned light
can be used to trap and guide atoms outside  the fiber. We have shown that,  
when both the input light fields are of circular polarization, a set of 
trapping minima of the potential in the transverse plane is formed as a ring around the fiber. In this case,
the atoms can be confined to a  cylindrical \textit{shell} around the fiber. 
When one or both of the input light fields are of linear polarization, the potential achieves its minimum value  
at two single points in the transverse plane. In this case, the atoms can be confined along two straight \textit{lines} parallel to the fiber axis. 
Due to the thin thickness of the fiber, we can use far-off-resonance fields with substantially differing evanescent decay lengths to produce a net potential with  a \textit{large} depth,  a \textit{large} coherence time, and  a \textit{large} trap lifetime. For example, a 0.2-$\mu$m-radius silica fiber carrying 30 mW of
1.06-$\mu$m-wavelength  light   and 29 mW of 700-nm-wavelength  light, both fields  are circularly polarized at the input, gives for cesium atoms a trap depth of 2.9 mK, a coherence time of 32 ms, 
and a recoil-heating-limited trap lifetime of 541 s.

\begin{acknowledgments}
This work was carried out under the 21st Century COE program on ``Coherent Optical Science''.
\end{acknowledgments}

\end{document}